%% file: rad.tex
\documentclass[acmsmall]{acmart}
\usepackage{lineno}
\input{mydef}

\AtBeginDocument{%
  \providecommand\BibTeX{{%
    \normalfont B\kern-0.5em{\scshape i\kern-0.25em b}\kern-0.8em\TeX}}}

\setcopyright{acmcopyright}
\copyrightyear{2021}
\acmYear{2021}
\acmDOI{}

\acmJournal{JACM}
\acmVolume{--}
\acmNumber{--}
\acmArticle{---}
\acmMonth{3}

\begin{document}

%%%%%%%%%%%---SETME-----%%%%%%%%%%%%%
\title{Entropy Maximization in Sparse Matrix by Vector Multiplication ($\max_E SpMV$) }

\author{Abhishek Jain}
\email{abhishek@xilinx.com}
\author{Ismail Bustany}
\email{ismailb@xilinx.com}
\author{Henri Fraisse}
\email{henrif@xilinx.com}
\author{Mansimran Benipal}
\email{mansimra@xilinx.com}
\author{Dinesh Gaitonde}
\email{dineshg@xilinx.com}
\author{Paolo D'Alberto}
\email{paolod@xilinx.com}
\affiliation{%
  \institution{Xilinx}
  \streetaddress{2100 Logic Dr}
  \city{San Jose}
  \state{California}
  \postcode{95124}
}

\renewcommand{\shortauthors}{Jain et al.}

\begin{abstract}
The peak performance of any SpMV depends primarily on the available
memory bandwidth and its effective use. GPUs, ASICs, and new FPGAs
have higher and higher bandwidth; however, for large scale and highly
sparse matrices, SpMV is still a hard problem because of its random
access pattern and workload imbalance.  Here, we show how to turn
randomness to our advantage. We propose a matrix permutation
pre-processing step that aims to maximize the entropy of the
distribution of the nonzero elements.  We seek any permutation that
uniformly distributes the non-zero elements' distribution, thereby
generating a SpMV problem that is amenable to work load balancing or
to speed up sort algorithms.  We conjecture these permutations would
be most effective for matrices with no dense rows or columns and, as
in preconditioning, when the matrix is reused. We shall show that
entropy maximization is an optimization that any architecture may take
advantage although in different ways. Most importantly, any developer
can consider and deploy. We shall present cases where we can improve
performance by 15\% on AMD-based (GPU-CPU) systems.

\end{abstract}

\maketitle

\section{Introduction} 
\label{sec:introduction}

To define the scope of our work, the obvious questions to ask are:
first, what randomization or entropy maximization is in the context of
sparse matrices; second, why would we use it; third, when it does
work.  We shall provide formal definitions in the following
sections. Here briefly, we will permute randomly the rows and columns
of a sparse matrix before multiplying it by a dense vector (SpMV) with
the aim of speeding this operation (accordingly the input and output
vectors will be permuted).
\begin{equation}
(P_r\Vc{y}) =(P_r \Vc{A} P_c)* (P_c^{-1}\Vc{x}) 
\end{equation}

Undoubtedly, this scheme requires some restrictions about the matrix
structure: for example, it must have no or few dense columns or rows.
In this unfortunate case, a sparse/dense partitioning scheme should be
used and different algorithms/hardware could be deployed separately
instead. Here suffice to say that the partitioning advantages rely on
a clear definition of density.  For the remainder of this manuscript,
we shall assume the former nonzero structure. We use randomization
because it is the poor man's way for preconditioning SpMV in our
context, and we do not mean it in a pejorative sense.

Preconditioning speeds up the convergence rate of an iterative linear
solver by linearly transforming the associated matrix into a form that
affords a faster reduction of the residual error at every iteration.
The cost of this transformation is justified by the runtime reduction
it affords.  Likewise, we foresee randomization playing a similar role
for SpMV in the context of iterative linear solvers and other methods
(e.g in convolutions) where the matrix is reused.

Sparse linear algebra and GraphBLAS kernels are memory bound and there
is a common thread in the scientific computing community to develop
acceleration libraries mostly for multi-core systems. These
predominantly include multi-core processors and GPUs. The goal is
a balanced work distribution and, when applicable, minimal
communication
\cite{DBLP:journals/siamsc/KayaaslanAU18,DBLP:conf/ieeehpcs/PageK18}. When
storage strategy and algorithms must be considered together then GPUs
provide the work horse for abundant thrust in research
\cite{DBLP:journals/topc/AnztCYDFNTTW20}. These works aim at optimal
solutions and strive for a clear and complete
understanding/exploitation of the software-hardware interface; usually
the hardware is composed of symmetric computational
units. Interestingly, the SpMV's space and time complexity, which are
small, may not warrant more performance because we typically end up
utilizing only one-thousandth fraction of the available hardware
capacity.

The peak performance of any SpMV accelerator depends primarily on the
available memory bandwidth (i.e., DRAM such as DDR or HBM) and the
capability of the accelerator to effectively use it.  Because SpMV is
memory-bound, a more important metric than peak performance alone is
the fraction of bandwidth utilized, which captures the overall
efficiency of the architecture. GPU platforms exhibit very high
bandwidth, see the experimental Section \ref{sec:experimentalresults}:
Ellesmere DDR5 224GB/s, Fiji HBM 512GB/s, and Vega 20 HBM 1TB/s.
Although utilizing this much bandwidth efficiently is difficult for
large scale and highly sparse matrices due to very high random access
pattern.  Custom architectures based on FPGA or ASIC devices can
maximize bandwidth utilization by highly customized data-paths and
memory hierarchy
designs~\cite{,fowers2014high,grigoras2016cask,zhou2019hitgraph}. Most
of the existing accelerators saturate the relatively low memory
bandwidth available on FPGA platforms (less than 80
GB/s)~\cite{nagar2011sparse,townsend2013reduce,fowers2014high,grigoras2016cask,li2016data,zhou2019hitgraph}.
Modern FPGA platforms have multiple HBM stacks to provide large memory
bandwidth. However, there is no implementation (currently available)
that saturates all of the available DRAM bandwidth for SpMV kernel on
HBM-enabled FPGA platforms. Scalability of accelerator design remains
a major concern, and it is an active area of research.
%\newline

FPGA platforms used in early works exhibit low peak performance due to
the scarcity of external memory
bandwidth~\cite{kestur2012towards,zhuo2005sparse,fowers2014high}.  For
example, Microsoft's implementation of SpMV uses an FPGA platform
which only has 2 DDR2-400 memory banks with a resulting bandwidth of
6.4 GB/s~\cite{kestur2012towards}.  The accelerator is running at 100
MHz, it reads 64 Bytes of data every cycle, which corresponds to 5
non-zeros at every cycle (a non-zero is about 12 Bytes). At best, the
peak performance is 10 double precision operations every cycle at 100
MHz, which is 1 GFLOPS (only). In 2009, Convey systems Inc. released
the Convey HC-1 FPGA platform.  It has 16 DDR2-677 memories resulting
in overall 80 GB/s memory bandwidth~\cite{nagar2011sparse}. The
accelerator logic runs at 150 MHz.  It consumes 512 Bytes of data
every cycle, which corresponds to around 40 non-zeros every cycle.  At
best, the peak performance is 80 double precision operations every
cycle at 150 MHz, which is 12 GFLOPS.
%\newline

One of the key building blocks for custom architecture solutions is a
multi-ported buffer used to storing vector
entries~\cite{fowers2014high}.  During execution, multiple column
indices are used as addresses to read corresponding vector entries; we
shall provide more details about the application in Section
\ref{sec:notations}.  Designing a buffer with a very large number of
read ports is challenging.  One solution is {\em banking} as a
mechanism to store partitioned vector entries.  Although banking could
allow very high throughput indexing unless the same entry is required
multiple times and its reads are purely sequential causing loss of
bandwidth.  For example, hashing techniques and data duplication are
possible solutions for this problem. However, another issue arises:
When we distribute SpMV computations across $p$-nodes, some of the
nodes, say $k$, finish later than the rest because of unbalanced work
loads (i.e., number of nonzero element) in row/column major
traversal. This is a common phenomena for matrices where few rows or
columns are dense. These $k$ nodes are referred to as {\em laggard
  nodes}.  By applying random permutation of columns/rows, we are
attempting to balance the loads across all $p$ workers so that there
are no laggards. From this hardware vantage point, randomization or
maximizing the entropy of the non-zero element distribution is an
optimization transform and provides a clear context for our work.
%\newline

Clearly, accelerating SpMV is a hard many-parameters optimization
problem dependent on the choice of algorithm, data structures, and
dedicated hardware (CPU, GPUs, FPGA's, Custom ASIC's).  Rather, our
goal is to provide a tool, we may say a naive tool, to help understand
how the structure of the matrix may affect the HW-SW solution.  For
the readers in the field of algorithms, SpMV can be mapped into a
sorting algorithm. For example, finding elements $x_{i,j}$ and
$x_{i,k>j}$ in a sparse matrix requires to find row $i$ and then
columns $j$ and $k$.  Sorting is a method to find if an element is in
a list with no prior or limited knowledge of its contents.  Sorting
can be used to prepare the matrix and to find elements in between
sparse matrices and sparse vectors. In custom architectures, sorting
networks are used to route matrix and vector elements to functional
units. In a sense, if one is stuck with a sorting algorithm and a poor
distribution, randomization may alter the distribution and throttle
performance. Interestingly, the best sorting algorithm is a function
of the distribution of the elements \cite{LiGP2004,HuangGl2009}.

We organize this work as follows: In Section \ref{sec:notations}, we
define the matrix by vector operation; in Section
\ref{sec:randomization}, we define what we mean by randomization or
entropy maximization. We use randomization to create a uniform
distribution in Section \ref{sec:uniform} and measure uniformity
%by nothing else than
by entropy in Section \ref{sec:entropy}. We present how we drive our
experiments to show the effects of randomization in Section
\ref{sec:measuring}. In the last sections, we present a summary of the
results: we present our work loads for the given benchmarks in Section
\ref{sec:workload}, and the complete set of measures for an AMD CPU
and GPUs systems in Section \ref{sec:experimentalresults} .

\section{Basic Notations}
\label{sec:notations}
Let us start by describing the basic notations so we can clear the
obvious (or not).  A Sparse-matrix by vector multiplication {\em SpMV}
on an (semi) ring based on the operations $(+,*)$ is defined as
$\Vc{y} = \M \Vc{x}$ so that $y_i = \sum_j M_{i,j}*y_j$ where $M_{i,j}
\eq 0$ are not represented nor stored. Most of the experimental
results in Section \ref{sec:experimentalresults} are based on the
classic addition (+) and multiplication (*) in floating point
precision using 64 bits (i.e., double floating point precision) albeit
are extensible to other semi-rings.  For instance, it is well known
that SpMV defined on the semi-ring (min,+) is a kernel in computing an
all-pairs shortest paths starting with a graph adjacency matrix, and
in using a Boolean algebra we can check if two nodes are connected,
which is slightly simpler.
%\newline

We identify a sparse matrix $\M$ of size $M\times N$ as having
$O(M+N)$ non-zero elements, number of non zero {\em nnz}. Thus the
complexity of $\M \Vc{x}$ is $O(M+N) \approx 2nnz$. Also, we must read
at least $nnz$ elements and thus the complexity is $\Theta(M+N)
\approx nnz$. We can appreciate that reading the data is as complex as
the overall operation. Of course, the definition of sparsity may
vary. We represent the matrix $\M$ by using the coordinate list {\em
  COO} or and the compressed sparse row {\em
  CSR}\footnote{a.k.a. Compressed row storage {\rm CRS}.}
formats. The COO represents the non-zero of a matrix by a triplet
$(i,j,v)$; very often there are three identical-in-size vectors for
the ROW, COLUMN, and VALUE. The COO format takes $3\times nnz$ space
and two consecutive elements in the value array are not bound to be
neither in the same row nor column. In fact, we know only that
$VALUE[i] = M_{ROW[i],COLUMN[i]}$.

The CSR format stores elements in the same row and with increasing
column values consecutively. There are three arrays V, COL, and
ROW. The ROW is sorted in increasing order.  Its size is $M$, and
$ROW[i]$ is an index in V and COL describing where $i$-th row starts
(i.e., if row $i$ exists).  Accordingly, $M_{i,*}$ is stored in
$V[ROW[i]:ROW[i+1]]$. The column indices are stored at
$COL[ROW[i]:ROW[i+1]]$ and sorted increasingly. The CSR format takes
$2\times nnz + M$ space and a row vector of the matrix can be found in
$O(1)$.
%\newline

The computation  $y_i = \sum_j M_{i,j}*x_j$ is a sequence of scalar
products and, using the CSR format, is computed as follows:

\[ Index = ROW[i]:ROW[i+1] \]
\[
y_i =  \sum_{\ell\in Index} V[\ell] * x_{COL[\ell]}  
\]
%\newline

The matrix row is contiguous (in memory) and rows are stored in
increasing order. However, the access of the dense vector $\Vc{x}$ has
no particular pattern, well increasing.
%\newline

The COO format can be endowed with certain properties. For example, we
can sort the array by row and add row information to achieve the same
properties of CSR. In contrast, transposing a "sorted" COO matrix
simply entails swapping of the arrays ROW and COL. Think about matrix
multiply (one of us does constantly).  Each scalar product achieves
peak performance if the reads of the vector $\Vc{x}$ are streamlined
as much as possible and so the reads of the vector $V$. If we have
multiple cores, each could compute a subset of the $y_i$ and a clean
data load balancing can go a long way. If we have few functional
units, we would like to have a constant stream of independent $*$ and
$+$ operations but with data already in registers. That is, data
pre-fetch will go a long way especially for $x_{COL[i]}$, which may
have an irregular pattern.

\section{Randomization and Entropy Maximization}
\label{sec:randomization}
We define {\em Randomization} as row or column permutation transform
of the matrix $\M$ (thus a permutation of $\Vc{y}$ and $\Vc{x}$), and
we choose these by a pseudo-random process. The obvious question is to
as why should we seek randomization transforms?  The sparsity of
a given matrix $\M$ has a non-zero element distribution induced by the
nature of the original problem or by some imposed ordering on the
respective nodes of its associated graph.  This distribution may be
computationally incompatible with the chosen algorithm or
architecture. For instance, it can induce some load imbalance in the
computation.  We could break this load imbalance by seeking to
maximize entropy for this distribution. Our conjecture is that would
favor the average case performance rather than the worse case when
operating on the "max-entropy transformed" matrix.

For linear system solvers, if we know the matrix $\M$, and we know the
architecture, preconditioning (when affordable) is a better solution.
If we run experiments long enough, we choose the best permutation(s)
for the architecture, permute $\M$, and go on testing the next.  On
one end, preconditioning exerts a full understanding of both the
matrix (the problem) and how the final solution will be computed
(architecture).  On the other end, the simplicity of a random
permutation requires no information about the matrix, the vector, and
the architecture. Such a simplicity can be exploited directly in
Hardware. We are after an understanding when randomization is just
enough: We seek to let the hardware do its best with the least effort,
or at least with the appearance to be effortless.
%Also we shall show there are different flavors of randomization.

Interestingly, this work stems from a sincere surprise about
randomization efficacy and its application on custom SpMV. Here, we
wish to study this problem systematically so that to help future
hardware designs. Intuitively, if we can achieve a uniform
distribution of the rows of matrix $\M$ we can have provable
expectation of its load balancing across multiple cores. If we have a
uniform distribution of accesses on $\Vc{x}$ we could exploit column
load balancing and exploit better sorting algorithms: In practice, the
reading of $\Vc{x}_{COL[i]}$ can be reduced to a sorting, and there we know
that different sparsity  may require different algorithms. This may be 
a lot to unpack but it translates to a better performance of the
sequential algorithm without changing the algorithm or to improved bandwidth
utilization.

We will show that (different) randomness affects architectures and
algorithms differently, making randomization a suitable optimization
transform especially when the application and hardware are at odds:
Hardware (unless programmable) is difficult to change and the matrix
sparsity is simple to change. We want to show that there is a
randomness hierarchy that we can distinguish as global and
local. There are simple-to-find cases where the sparsity breaks
randomness optimization.  For instance, matrices with dense rows or
columns are better partitioned into sparse and dense components and
operated on separately.
%We want to show that this study uses common tool, open
%software tools and sometimes naive experiments; however, we can infer
%properties applicable to proprietary and custom solutions.
%Hi Paolo, above two sentences need further clarification
%\newpage 
\Doublefigure{.49}{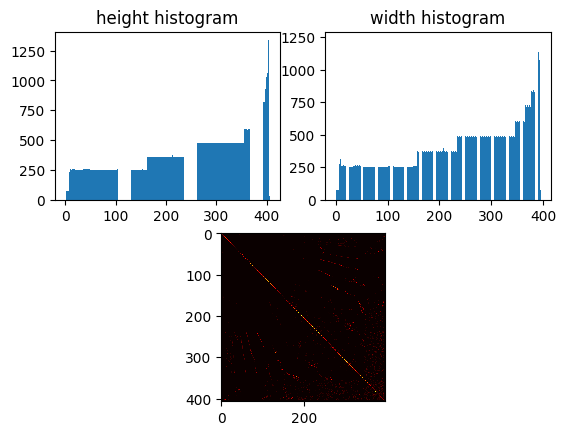}{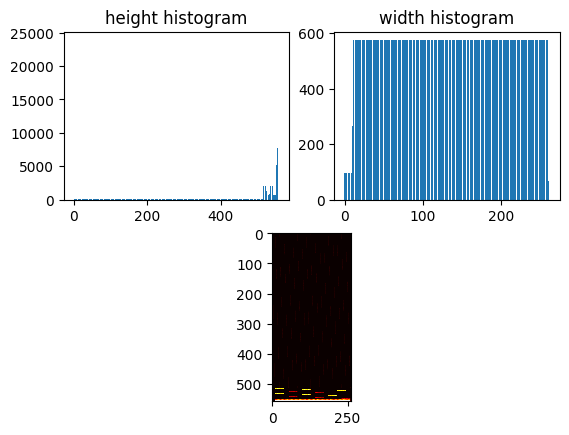}{Left:
  OPF 3754. Right: LP OSA 07. These are histograms where we represent
  normalized buckets and counts}{fig:one}

\section{Entropy}
\label{sec:entropy}
Patterns in sparse matrices are often visually pleasing, see Figure
\ref{fig:one} where we present the height histogram, the width
histograms, and a two-dimensional histogram as heat map. We will let
someone else using AI picture classification. Intuitively, we would
like to express a measure of uniform distribution and here we apply
the basics: {\em Entropy}. Given an histogram $i\in[0,M-1]$ $h_i \in
\N$, we define $S =\sum_{i=0}^{M-1}h_i$ and thus we have a probability
distribution function $p_i = \frac{h_i}{S}$. The {\em information} of
bin $i$ is defined as $I(i) = -\log_2 p_i$. If we say that the
stochastic variable $X$ has PDF $p_i$ than the entropy of $X$ is
defined as.

\begin{equation}
  \label{eq:entropy}
  H(x) = -\sum_{i=0}^{M-1} p_i\log_2p_i = \sum_{i=0}^{M-1}p_i I(i) =
  E[I_x]
\end{equation}
The maximum entropy is when $\forall i, p_i = p = \frac{1}{M}$; that
is, we are observing a uniform distributed event.  Our
randomization should aim at higher entropy numbers. The entropy for
matrix LP OSA 07 is 8.41 and for OPF 3754 is 8.39. We use the entropy
specified in the Scipy stats module.  A single number is concise and
satisfying. If you are pondering why they are so close contrary to
their sparsity we discuss this next.

\section{Uniform distribution}
\label{sec:uniform}
We know that we should {\bf not} compare the entropy numbers of two
matrices because entropy does not use any information about the order
of the buckets, it uses only their probabilities. By construction, the
matrices are quite different in sparsity and in shapes, however their
entropy numbers are close.  Two matrices with the same number of
non-zeros, spaced well enough in the proper number of bin, will have
the same entropy. To appreciate their different sparsity, we should
compare their entropy distributions by Jensen-Shannon measure
\cite{dalberto2012nonparametric} or we could use cumulative
distribution function (CDF) measures, which imply an order. Here, we
use a representation of a hierarchical 2D-entropy, see Figure
\ref{fig:two}, where the entropy is split into 2x2, 4x4 and 8x8 (or
fewer if the distribution is not square). We have hierarchical entropy
heat maps.

\doublefigure{.40}{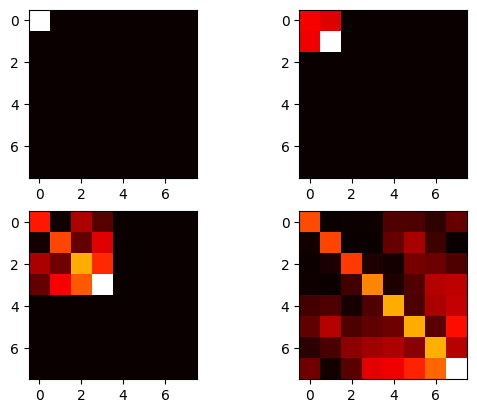}{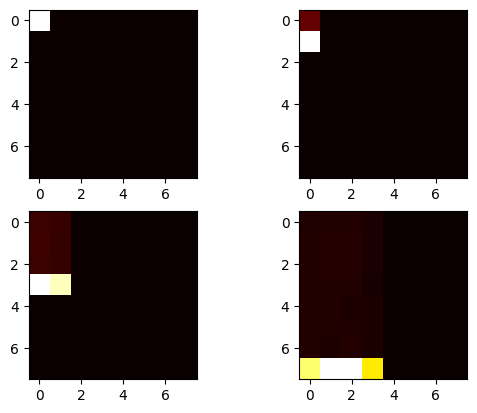}{Hierarchical 2D entropy for OPF 3754 (left) and LP OSA 07 (right). }{fig:two}

We can see that even a small 2D-entropy matrix summarizes the nature
of the original matrix because it has spatial information. In this
work, the entropy matrix is used mostly for visualization purpose more
than for comparison purpose. Of course, we can appreciate how the
matrix LP OSA 07 has a few very heavy rows and they are
clustered. This matrix will help us showing how randomization need
some tips. Now we apply row and column random permutation once by row
and one by column: Figure \ref{fig:three}: OPF has now entropy 11.27
and LP 9.26. The numerical difference is significant. The good news is
that for entropy, being an expectation, we can use simple techniques
like bootstrap to show that the difference is significant or we have
shown that Jensen-Shannon can be used and a significance level is
available. What we like to see is the the hierarchical entropy heat
map is becoming {\em more} uniform for at least one of the matrix.

\doublefigure{.40}{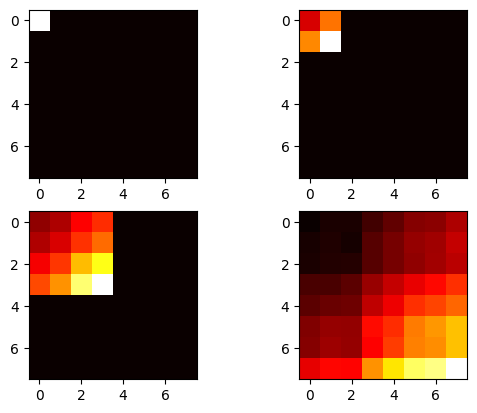}{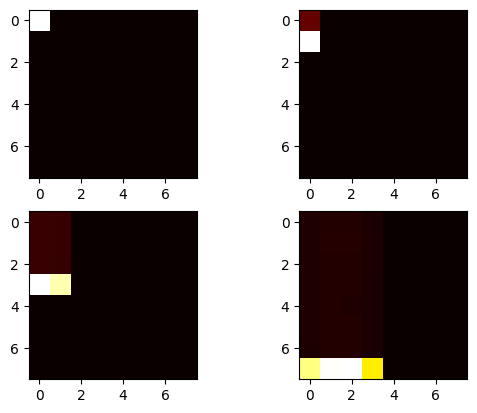}{Hierarchical 2D entropy after row and column random permutation for OPF 3754 (left) and LP OSA 07 (right). }{fig:three}

In practice, permutations need some help especially for relatively
large matrices. As you can see, the permutation affects locally the
matrix. Of course, it depends on the implementation of the random
permutation, we use {\em numpy} for this. It is reasonable that a
slightly modified version of the original is still a random selection
and unfortunately they seem too likely in practice. We need to
compensate or help the randomization. If we are able to identify the
row and column that divide high and low density, we could use them as
pivot for a shuffle like in a quick-sort algorithm. We could apply a
sorting algorithm but its complexity will the same of SpMV. We use a
gradients operations to choose the element with maximum steepness,
Figure \ref{fig:four} and \ref{fig:five}.

LP achieves entropy 8.67 and 9.58 and OPF achieves 10.47 and 11.40.

\doublefigure{.40}{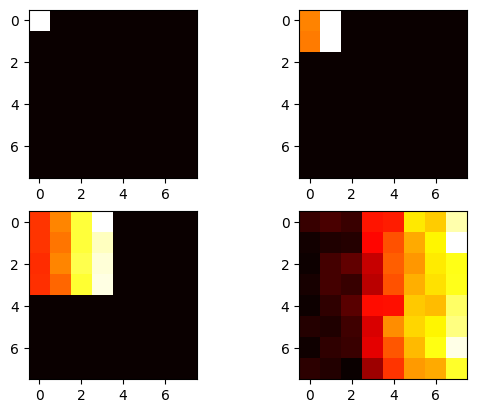}{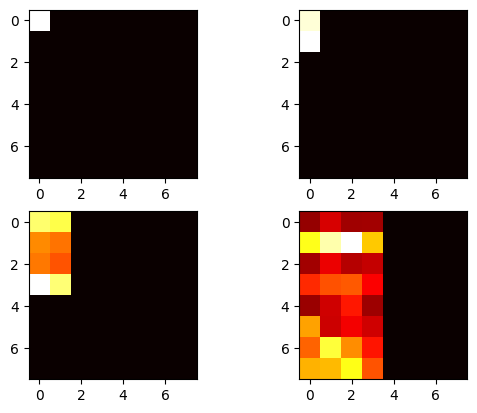}{Hierarchical 2D entropy after height gradient based shuffle and row random permutation for OPF 3754 (left) and LP OSA 07 (right). }{fig:four}

\doublefigure{.40}{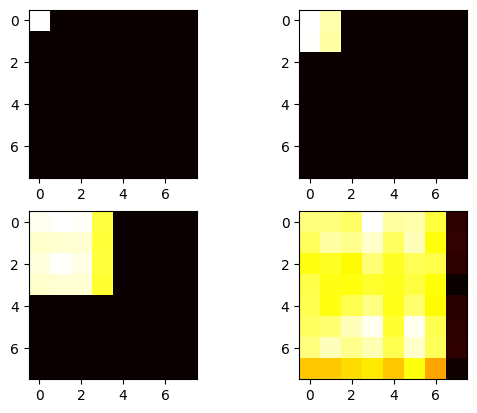}{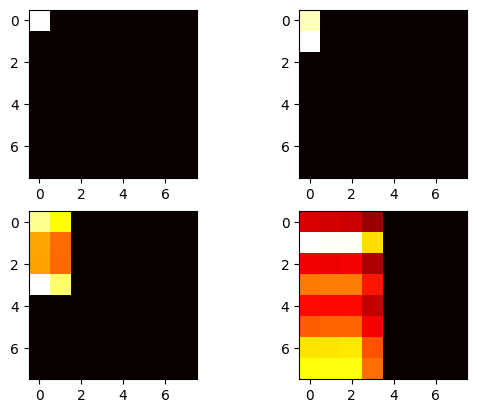}{Hierarchical 2D entropy after height and width gradient shuffle and row and column random permutation for OPF 3754 (left) and LP OSA 07 (right). }{fig:five}

If the goal is to achieve a uniformly sparse matrix, it seems that we
have the tools to compute and to measure such a sparsity. We admit
that we do not try to find the best permutation. But our real goal is
to create a work bench where randomization can be tested on different
architectures and different algorithms. A randomization with a
measurable uniform distribution is preferable than just random. We are
interested to find out when random is enough or not enough. Also,
consider that to achieve a uniform distribution, we do not need a
random transformation and any permutation balancing the number of
non-zero is possible, but for now not looked for.

\section{Measuring the randomization effects}
\label{sec:measuring}

Whether or not this ever applied to the reader, when we have timed
algorithms (i.e., measure execution time), we came to expect
variation.  The introduction of randomization may hide behind the ever
present variance, after all these are algorithms on {\em small}
inputs: small error can be comparable to the overall execution
time. Here, we must address this concern even before describing the
experiments.

First, we execute every algorithm between 1000 and 5000 times. The
time of each experiment is in the seconds, providing a granularity for
which we are confident the measuring time error is under
control. Thus, for each experiment we provide an average execution
time: we measure the time and we divide by the number of trials. Cold
starts, the first iteration, are still accounted. To make the measure
portable across platform we present GFLOPS, that is, Giga ($10^{12}$)
floating operations per second: $2*nnz$ divided by the average time in
seconds. 

Then we repeat the same experiment 32 times. Permutations in {\em
  numpy} Python uses a seed that is time sensitive: thus every
experiment is independent from the previous. The number 32 is an old
statistic trick and it is a minimum number of independent trials to
approximate a normal distribution. In practice, they are not but the
number is sufficient for most of the cases and it is an excellent
starting point.

A short hand legend: {\bf Reg} is the regular matrix without any
permutation; {\bf R} stands for random {\em Row} permutation; {\bf
  G-R} stands for gradient-based row shuffle and random row
permutation; {\bf G-C} stands for gradient-based column shuffle and
random column permutation; {\bf R-C} stands for random row and column
permutation.  This legend is used in the pictures to be concise, in
the tables in the following sections, we use a verbose description. We
shall clarify the gradient based approach in the experimental results
section \ref{sec:experimentalresults}. Intuitively, we help the random
permutation by a quick targeting of high and low volume of the
histogram (and thus the matrix).

In Figure \ref{fig:five1}, we show two plots respectively of the CPU
performance using COO and CSR SpMV algorithms for the matrix OPF
3754. The figure represents histograms: The $x$ is GFLOPS and the $y$
label is the number of counts. Thus we show what is the performance
distribution of an algorithm.  We can see that the CSR algorithms are
consistent and the Regular (i.e., the original) has always the best
performance. Also the variance of the computation time is small and
the shape is approximately Gaussian.  Different story for the COO, the
permutations introduce long tails, thus $2\times$ performance
advantage.
\doublefigure{.45}{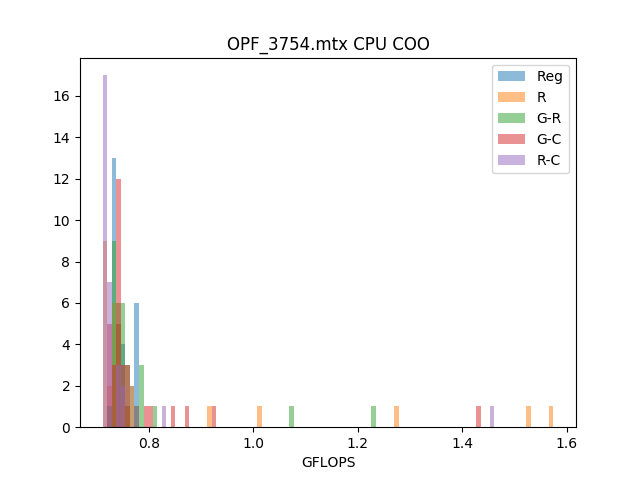}{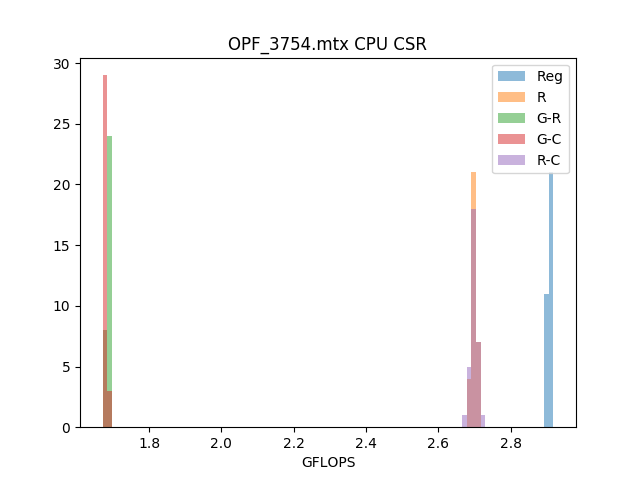}{CPU
  COO (left) and CPU CSR (left) for OPF 3754}{fig:five1}

\singlefigure{.45}{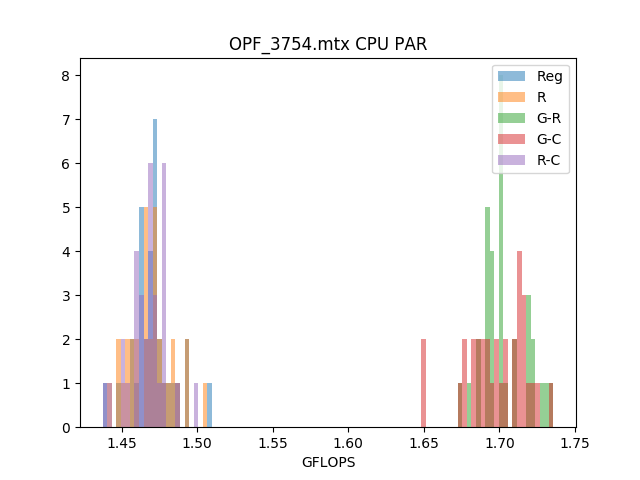}{ Parallel CPU CSR for OPF 3754}{fig:seven}
If we take the original matrix and split into parts having the same
number of rows, and execute them in parallel using different cores, we
can see in Figure \ref{fig:seven} that randomization is quite useful.

In Figure \ref{fig:six}, \ref{fig:six1} and \ref{fig:six2},
randomization is harmful to the GPU implementation. The OPF 375 matrix
is mostly diagonal, thus the vector $\Vc{x}$ is read in close
quarters, randomization breaks it.  If the load balance is fixed
(i.e., by dividing the matrix by row and in equal row), randomization
is beneficial.

\doublefigure{.45}{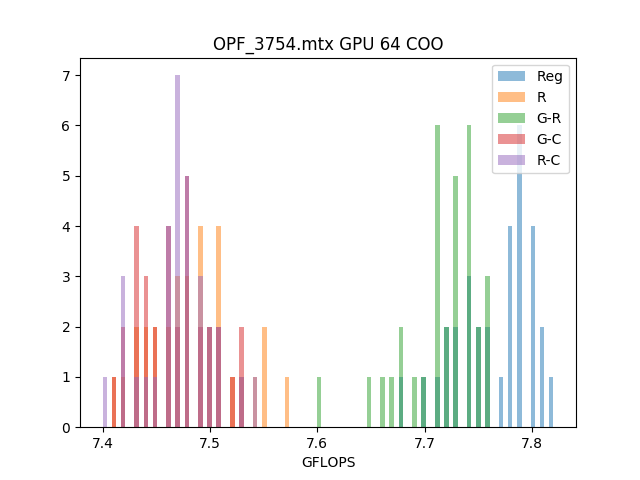}{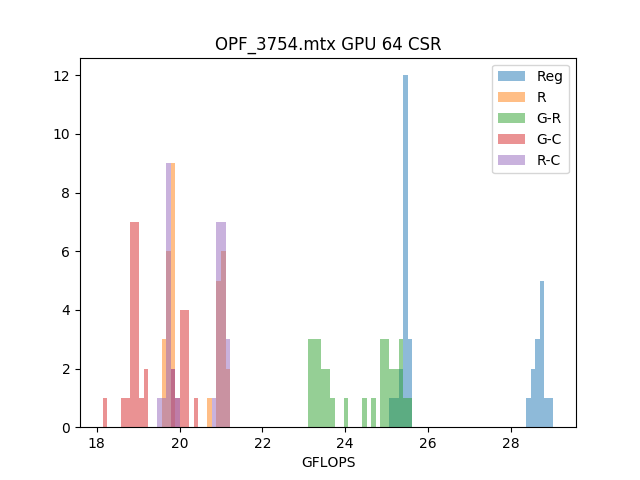}{Vega 20, GPU 64bits COO (left) and GPU CSR (right) for OPF 3754}{fig:six}
\doublefigure{.45}{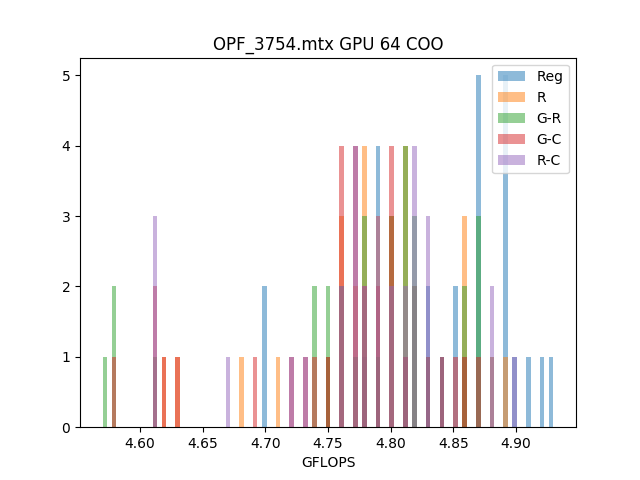}{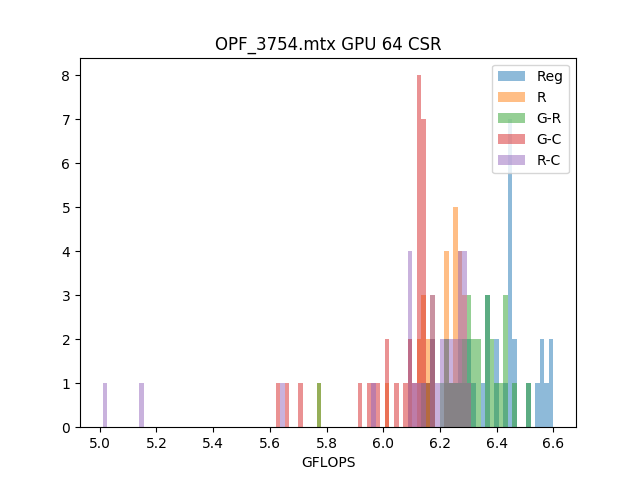}{Ellesmere, GPU 64bits COO (left) and GPU CSR (right) for OPF 3754}{fig:six1}
\doublefigure{.45}{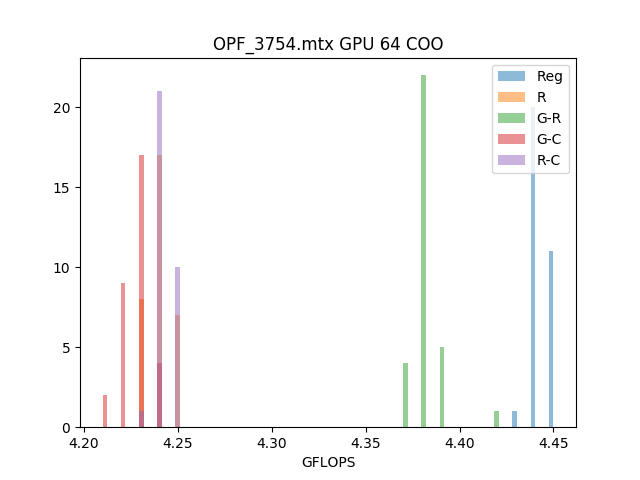}{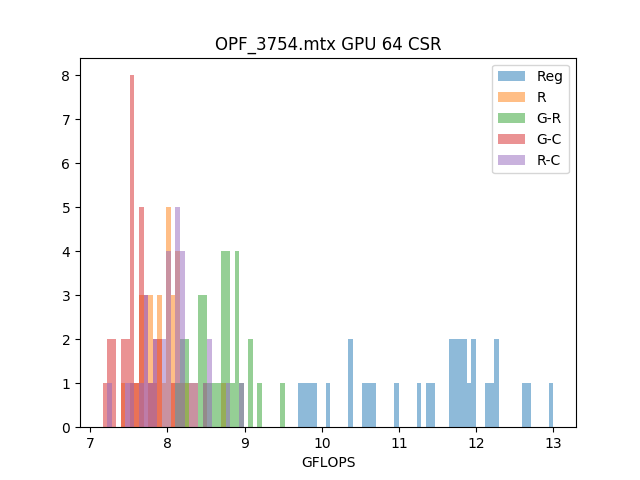}{Fiji, GPU 64bits COO (left) and GPU CSR (right) for OPF 3754}{fig:six2}

%\newpage

For matrix LP OSA 07, randomization helps clearly only for CPU CSR as
we show in Figure \ref{fig:eight}. In Figure \ref{fig:eight1},
\ref{fig:eight2}, and \ref{fig:eight3}, we can see that randomization
is harmful but for one GPU, we can show that a single exception is
possible (40\% improvement).

\singlefigure{.45}{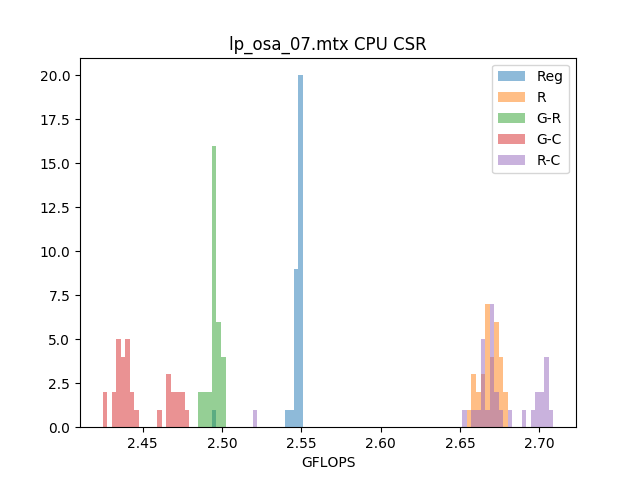}{   CPU CSR  for LP OSA 07}{fig:eight}

\doublefigure{.45}{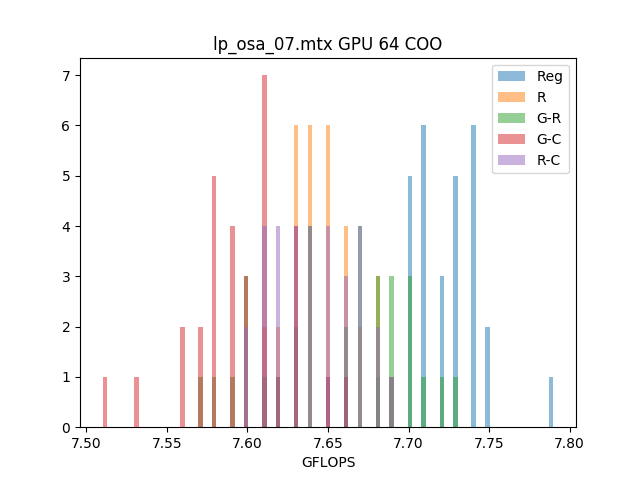}{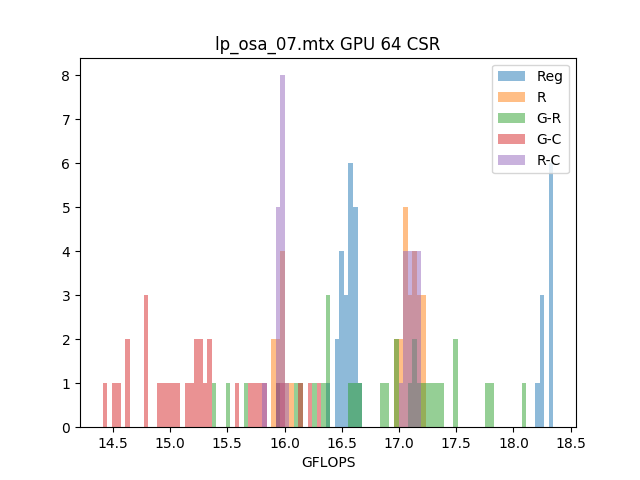}{Vega 20, GPU 64bits COO (left) and GPU CSR (right) for OPF 3754}{fig:eight1}
\doublefigure{.45}{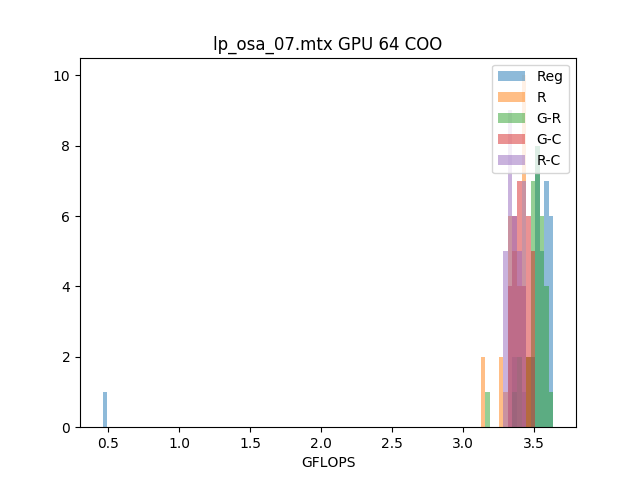}{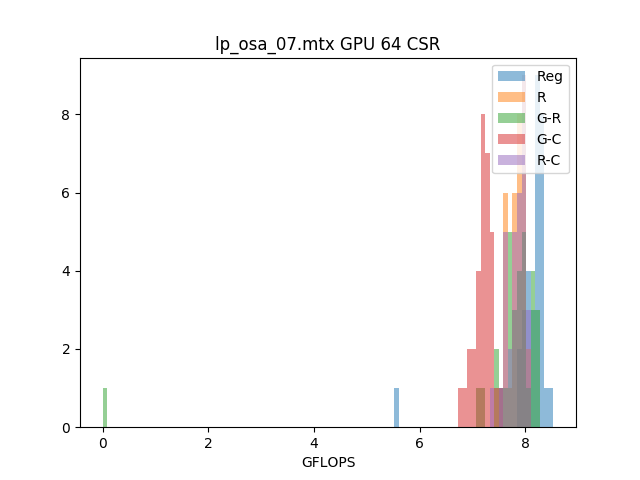}{Ellesmere, GPU 64bits COO (left) and GPU CSR (right) for OPF 3754}{fig:eight2}
\doublefigure{.45}{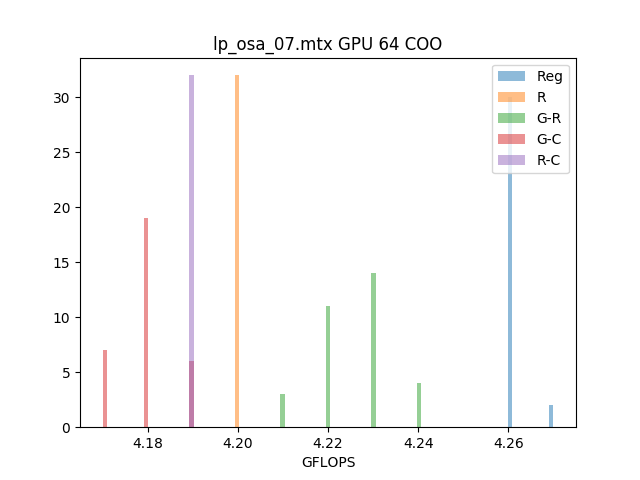}{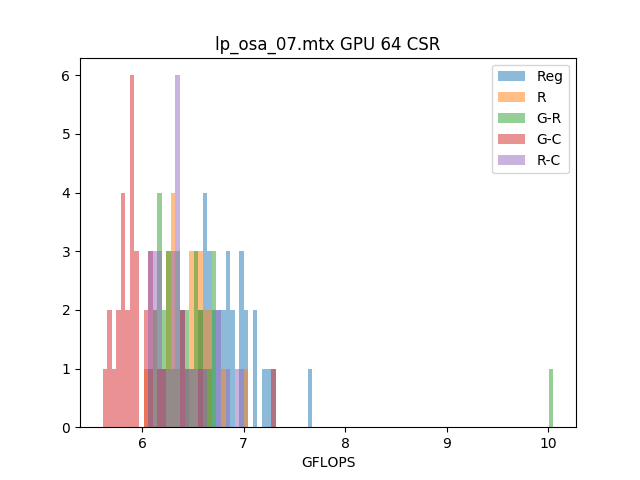}{Fiji, GPU 64bits COO (left) and GPU CSR (right) for OPF 3754}{fig:eight3}

%\newpage
An example, the matrix MULT DCOP 01, is where randomization is useful
for the CPU, GPU, and the parallel version Figure \ref{fig:9},
\ref{fig:10} - \ref{fig:11} and the gains can be up to
10-15\%. Consider, we can achieve these improvements without any
insights to the architecture, the algorithms and their relationships.

\doublefigure{.45}{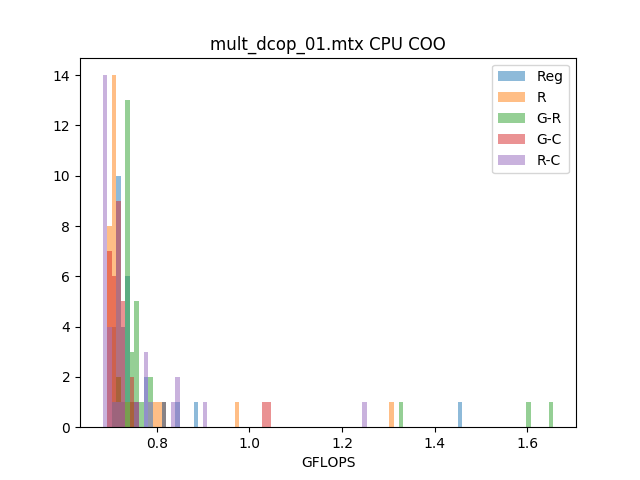}{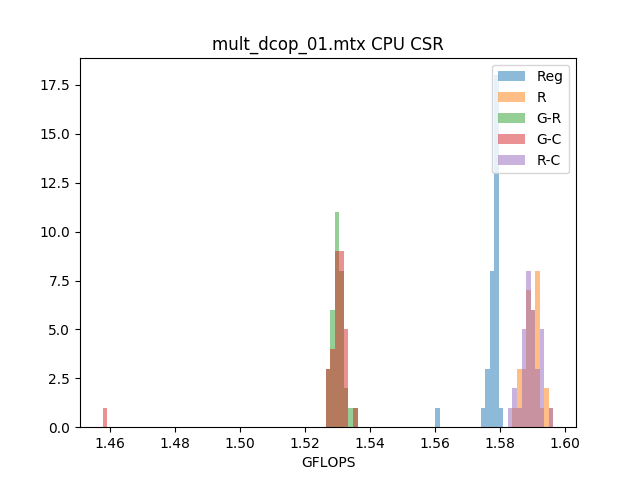}{CPU COO (left) and CPU CSR (right) for MULT DCOP 01}{fig:9}
\doublefigure{.45}{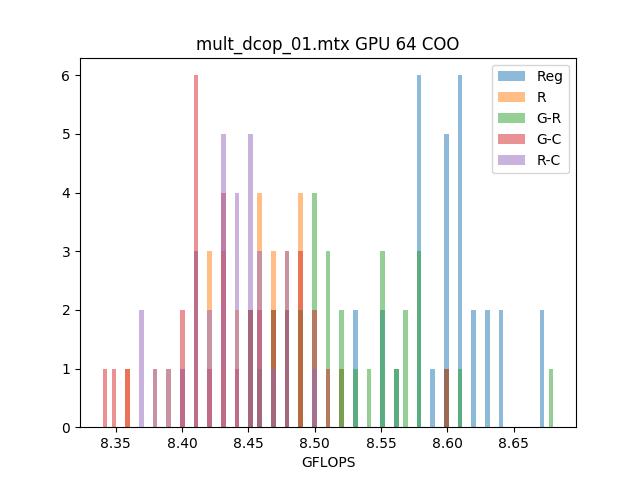}{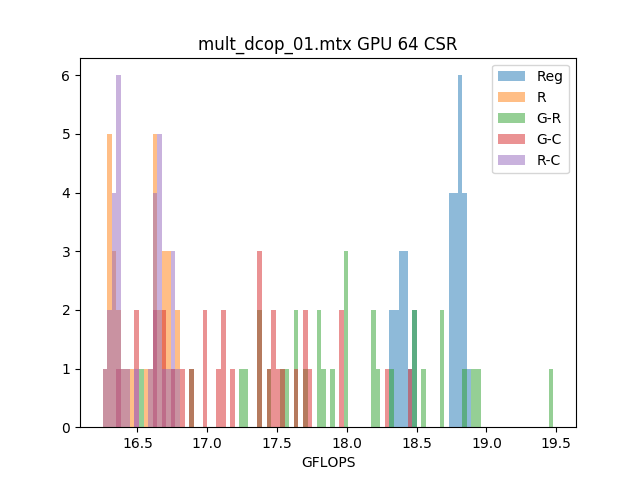}{Vega 20, GPU 64bits COO (left) and GPU CSR (right) for MULT DCOP 01}{fig:10}
\doublefigure{.45}{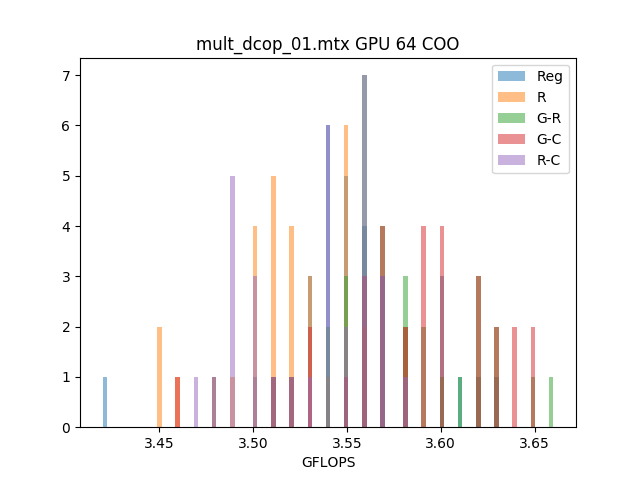}{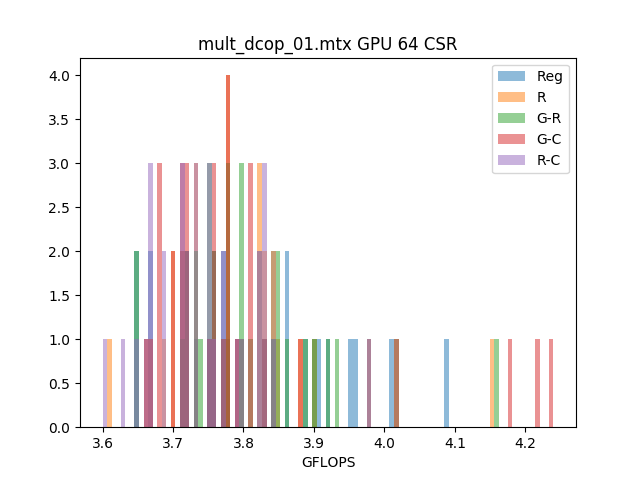}{Ellesmere,  GPU 64bits COO (left) and GPU CSR (right) for MULT DCOP 01}{fig:101}
\doublefigure{.45}{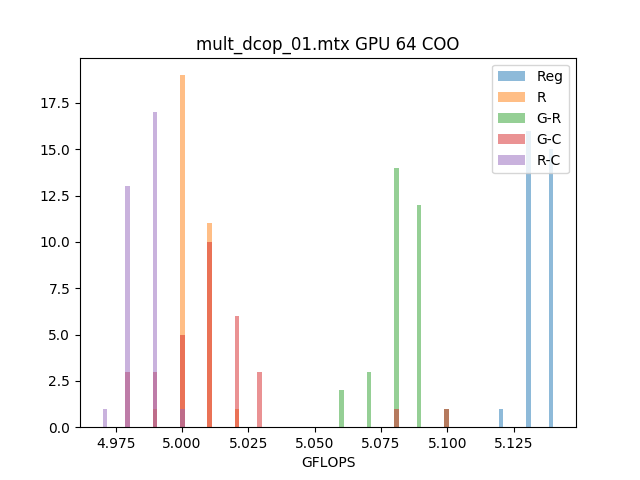}{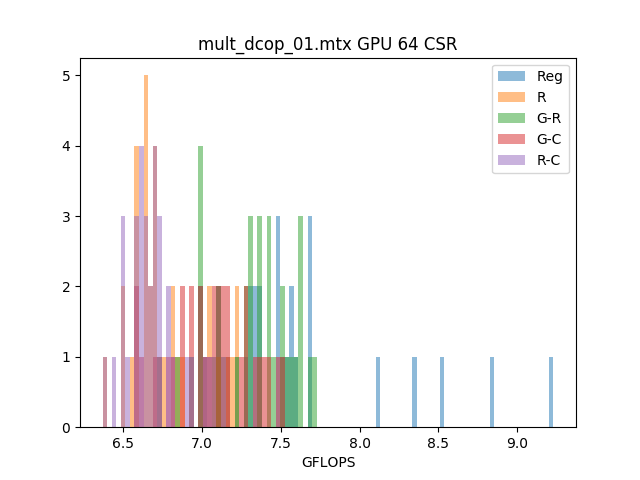}{Fiji, GPU 64bits COO (left) and GPU CSR (right) for MULT DCOP 01}{fig:102}
\singlefigure{.45}{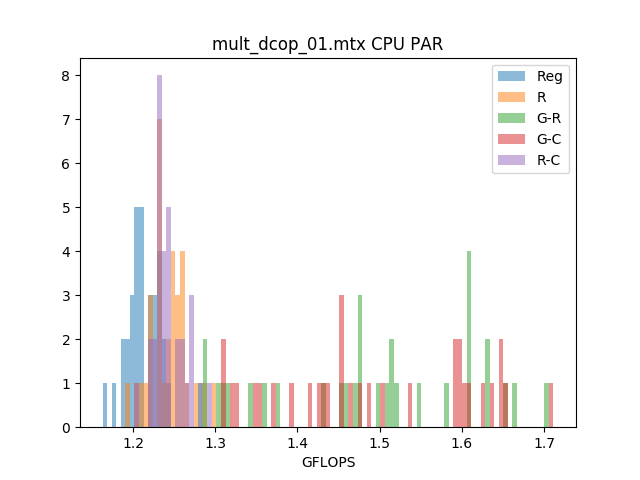}{ Parallel
  CPU CSR  for MULT DCOP 01 }{fig:11}

What does it mean when randomization does not work? The matrices we
use in this work are not chosen randomly (pun not intended), they are
the matrices that are difficult to handle in our custom SpMV engines
using a combination of sorting networks and systolic arrays. If
randomization does not work in our simplified work bench, will not
work in our specialized architecture because the reorganization of the
matrix or the input and output vector does not have the necessary
parallelism, data locality, and data streaming. We need to do
something else. In this case disrupting the memory pattern is not
sufficient. Thus, if we cannot beat the pattern, we must exploit it,
well not in this work.

\section{Workloads}
\label{sec:workload}

In the previous sections, we defined what we mean for randomization
and we present our tools of tricks for the measure of the effects of
randomization. Here we describe the work loads, the applications, we
use to test the effects of the randomization.

\subsection{Python COO and CSR algorithms}
\label{sec:pythoncoocsr}
The simplicity to compute the SpMV by the code $z = A*b$ in Python is
very rewarding. By change of the matrix storage format, $A =
A.tocsr(); z = A*b$, we have a different algorithm. The performance
exploitation is moved to the lower level.  The CSR implementation is
often two times faster but there are edge cases where the COO and COO
with randomization can go beyond and be surprisingly better: MUL DCOP
03 is an example where COO can do well.

Intuitively, Randomization can affect the performance because the
basic implementation is a sorting algorithm and it is a fixed
algorithm. There are many sorting algorithms and each can be optimal
for a different initial distribution. If we knew what is the sorting
algorithm we could tailor the input distribution. Here we just play
with it.

In Section \ref{sec:experimentalresults}, we present all the results
for CPU and GPUS. Keep in mind that these problems are hard, in the
sense they do not have fancy performance sheets (these architectures
can achieve Tera FLOPs sustained performance for dense computations).
If we go through diligently, we can see that there is a 15x
performance difference between the single thread CPU and Vega 20 GPU
(i.e, 3 vs 40 GFLOPS).

\subsection{Parallel CSR using up to 16 cores}
\label{sec:parcpu}
Python provides the concept of Pool to exploit a naive parallel
computation. We notice that work given to a Pool is split accordingly
to the number of elements to separate HW cores. We also noticed that
the work load move from a core to another, thus not ideal. Also we
notice that Pool introduce a noticeable overhead: a Pool of 1, never
achieves the performance of the single thread $z = A*b$. Using Pool
allows us to investigate how a naive row partitioning without counting
can scale up with number of cores. We tested by splitting the rows to
1--16 cores evenly (one thread per core) and we present the
performance for only the best configuration. The randomization goal is to
distribute the work uniformly: a balanced work distribution avoid the
unfortunate case where a single core does all the work. We are pleased
by the simplicity of the benchmark and we know we can do better.

\subsection{GPU COO and CSR algorithms}
\label{sec:gpucoocsr}
In this work, we use AMD GPUs and {\em rocSPARSE} is their current
software. The software has a few glitches but overall can be used for
different generation of AMD GPUs. We use the COO and CSR algorithms
and we provide performance measure for double precision only. The
ideas of using different GPUs: it is important to verify that the
randomization can be applied independently of the HW. We are not here
to compare performance across GPUs and CPUs. Often the limitation is
the software, how the software can exploit the hardware or how the
software will make easy to use a specific GPU. For example, the Fiji
architecture is clearly superior to the Ellesmere, however the latter
have better support and the system overall is more stable and user
friendly.

The performance of the CSR algorithm is about two times faster than the
COO. Most of the algorithms count the number of sparse elements in a
row and thus they can decide the work load partition
accordingly. Counting give you an edge but without changing the order
of the computation there could be cases where the work load is not
balanced and a little randomization could help and it does.

%\subsection{ FPGA ? (not necessary)}
\subsection{Randomization sometimes works}

For the majority of the cases we investigated and reported in the
following sections, Randomization does not work. However, there are
cases where randomization does work and does work for different
algorithms and architectures. If you are in the business of
preconditioning, permutations are pretty cheap. If you can find a good
one just consider like a preconditioning matrix, which it is. 

This shows also that HW has to be more conscious, well the HW designer
should, and accept that there are options at software level, at matrix
level and beyond.

%\section{Call for a different strategy}
%\label{sec:strategy}
%We want to find out randomization techniques that are suitable for
%custom hardware but also what are the most common and simple
%heuristics that can justified for any hardware.

\section{Experimental Results}
\label{sec:experimentalresults}
The main hardware setup is a AMD Threadripper with 16 cores. We have
three Radeon GPUs: Vega 20 7nm, Pro 2xFiji, and Pro 2xEllesmere.

Vega 20 can deliver 3.5TFLOPS in double precision and it has 1TB/s HBM
memory. Each Fiji provides 0.5 TFLOPS in double precision and has
512GB/s HBM, the card has two chips.  The Ellesmere provides 0.3TFLOPS
in double precision and has 224GB/s DDR5, the card has two chips. In
the performance plots presented earlier and in the following, you will
notice that the performance gap between these GPUs is not so
marked. We can safely state that $vega \sim 2\times Fiji$ and $Fiji \sim
2\times ellesmere$

There are 4 basic randomization formats:
\begin{itemize}
  \item {\bf Random Row Permutation}, we take the original matrix and
    permute the rows.
  \item {\bf Random Row and Column Permutation}, we take the original
    matrix and permute the rows and the columns.
  \item {\bf Gradient based row permutation}, we compute the row
    histogram and we compute the gradient: $h_{i+1} - h_i$. We find a
    single point where the gradient is maximum, this is the pivot for
    a shuffle like a magician would shuffle a deck of cards.  Then we
    permute the two parts randomly.
  \item {\bf Gradient based row and column permutation}, As above but
    also for the columns.
\end{itemize}

For large matrices (large number of columns and rows) a permutation
tends to be a close variation of the original, still a random
permutation. The gradient allows us to describe two areas of the
original matrix where there is a clear and de-marked density
variation: for example, there are two uniform distributed sub matrices
but one denser than the other. A shuffle redistributes every other
sample/card to different parts and these can be permuted locally.

We report in the following the performance results GFLOPS, we
introduce a $\boldsymbol *$ following the best performance. This is
tedious to read and, we assure, to write. The code and the results are
available as software repository. Remember each experiment is based on
32 different runs and thus we report maximum, minimum, and mean as a
summary. We use the symbol H for entropy.

\input{out.v.tex}
\input{out.elle.tex}
\input{out.fiji.tex}

%\input{conclusion.tex}

%%%%%%%%% -- BIB STYLE AND FILE -- %%%%%%%%
\bibliographystyle{ACM-Reference-Format} \bibliography{rad}
%%%%%%%%%%%%%%%%%%%%%%%%%%%%%%%%%%%%

%\appendix{Review and Response}
%\input{review.tex}
\end{document}

%% file: mydef.tex
\newcommand{\eq}[0]{{=}}

\newcommand{\N}{\mathbb{N}}

\newcommand{\M}{\mathbb{M}}

\newcommand{\Vc}[1]{{\boldsymbol #1}}

\newcommand{\doublefigure}[5]{{\begin{figure}[htb]%
\centering %
\includegraphics[width=#1\linewidth]{#2}
\includegraphics[width=#1\linewidth]{#3}
\caption{#4}%bio                 
\label{#5}%
\end{figure}}}

\newcommand{\Doublefigure}[5]{{\begin{figure*} %
\centering %
\includegraphics[width=#1\linewidth]{#2}%
\includegraphics[width=#1\linewidth]{#3}%
\caption{#4}%
\label{#5}%
\end{figure*}}}

\newcommand{\singlefigure}[4]{{\begin{figure}[htb] %
\centering %
\includegraphics[width=#1\linewidth]{#2}%
\caption{#3}%
\label{#4}%
\end{figure}}}

%% file: out.v.tex
\newpage
\twocolumn
\section{Vega VII and ThreadRipper}
\label{sec:vega}
{\tiny


}

%% file: out.elle.tex
\section{Ellesmere}
\label{sec:ellesmere}
{\tiny

%
}

%% file: out.fiji.tex
\section{Fiji}
\label{sec:fiji}
{\tiny

%
}